\documentclass[fleqn,twoside]{article}

\usepackage[cp866]{inputenc}
\usepackage[T2A]{fontenc}
\usepackage[russian,english]{babel}
\usepackage{amsfonts,amsmath,amssymb,amscd}
\usepackage{gc,epsf}

\newcommand{\beg}{\begin{equation}\label}
\newcommand{\eqnd}{\end{equation}}
\newcommand{\prt}{\partial}
\newcommand{\p}{\prime}
\newcommand{\bib}{\bibitem}

\heads{Vladimir~V.~Kassandrov}
      {General Solution of the Complex 4-Eikonal Equation...}

\begin{document}
\twocolumn[
\Arthead{8}{2002}{Suppl.2}{57}{62}
\Title{GENERAL SOLUTION OF THE COMPLEX 4-EIKONAL EQUATION\yy
	AND THE `ALGEBRODYNAMICAL' FIELD THEORY      }

\Author{Vladimir~V.~Kassandrov\foom 1 }         
{Department of General Physics, People's Friendship University of 
Russia,Ordjonikidze 3, 117419 Moscow, Russia}

\Abstract
    {We explicitly demonstrate the existence of 
twistor and ambitwistor structure for 4-dimensional complex eikonal equation 
(CEE) and present its general solution consisting of two different classes. 
For both, every solution can be obtained from a generating twistor 
function in a purely algebraic way, via the procedure similar to that used 
in Kerr theorem for shear-free null congruences. Bounded singularities of 
eikonal or its gradient define some particle-like objects with nontrivial 
characteristics and dynamics. Example of a new static solution to CEE with 
a ring-like singularity is presented, and general principles of algebraic 
field theory (`algebrodynamics') related to CEE are briefly discussed.}

] 

\email 1 {vkassan@sci.pfu.edu.ru}

\section{Ambitwistor structure of complex 4-eikonal equation}\label{ambitw}

The four-dimensional eikonal equation in flat Minkowski space-time $M$ 
(we choose the units where $c=1$)
\begin{equation}\label{eik}
\eta^{\mu\nu}\partial_\mu S \partial_\nu S = (\partial_t S)^2-\vert \vec
\nabla S \vert^2 = 0,
\end{equation}
is fundamental nonlinear Lorentz-invariant equation which governs the
propagation of wavefronts (discontinuties of fields) in every
relativistic theory ~\cite{fock,petrov}. Its complex generalization,
with $S(x,y,z,t)\in \mathbb{C}$, naturally arise in geometrical optics 
and quasiclassical limit of quantum theory ~\cite{maslov} as well as in 
GTR ~\cite{wilson}. Singular solutions of eikonal equation and its relations  
to other fields have been studied by H. Bateman ~\cite{bate} and,
recently, by E.T. Newman et al. ~\cite{newman}. Possible interpretation of
singularities as particles was discussed by Yu.P. Pyit'ev 
~\cite{pitiv} (for 5D-real eikonal) and A.M. Vinogradov ~\cite{vinogr} 
(from general viewpoint); see also our work ~\cite{trish}. 

In the version of biquaternionic analysis proposed in our works ~\cite{kass,
kass2,kass3,kass4} complex eikonal equation (CEE) (\ref{eik}) plays the role 
similar to that of 
the linear Laplace equation in complex analysis. Its nonlinearity arises as a 
direct consequence of the noncommutativity of biquaternion algebra. 

The method to obtain the {\it general integral} of eikonal equation, i.e. the 
solution dependent on an arbitrary function (starting from the so called 
"complete integral"), is well known (see, for example, ~\cite{landau,newman}). 
We not only extend these results to the complex holomorphic case. Making 
use of the obtained {\it twistor} and {\it ambitwistor} structure of CEE, we 
present its general solution, in a manifestly Lorentz and gauge invariant 
form, and demonstrate that arbitrary CEE' solution belongs to one of two and 
only two different classes. The procedure is extremely simple and can be 
regarded as generalization of 
the known {\it Kerr theorem} for shear-free null geodesics ~\cite{penrose, kerr}.  
The problem of classification and evolution of singularities of the eikonal 
wavefronts and of the related caustics also becomes quite transparent in the 
scheme developed below. 

In the $\mathbb{C}$-case {\it static}  solutions to CEE do 
exist, the known example is given by the Kerr twisting congruence with a 
ring-like singularity (leading to {\it stationary} Kerr metric in GTR). In 
section 2 we present another static solution to CEE with singularity of a 
similar ring-like structure. 

For one of the two fundamental classes of CEE' solutions deep connections 
with solutions of gauge and metric fields can be revealed. In particular, 
for Maxwell field related to these solutions of CEE {\it electric charge is 
necessarily a whole multiple of some "unit", elementary one.} These 
and other facts form the premises of the algebraic field theory -- 
{\it algebrodynamics} -- which we briefly discuss in section 3. 

To start with, let us represent CEE ~(\ref{eik}) in  equivalent Pfaffian 
form
\begin{equation}\label{pfaff}
dS = \varphi dX \psi,
\end{equation}
where the two $SL(2,\mathbb{C})$-spinors $\varphi=\{\varphi_A\}$ 
and $\psi=\{\psi_{A^\prime}\}$ 
are introduced, and $dX$ is the differential of hermitian matrix of space-time 
coordinates $X=\{X^{AA^\prime}\}, \\A,A^\prime,...=0,1$. For the latters we'll use 
the standard representation $X^{00^\prime}=u=t+z,~X^{11^\prime}=v=t-z,~
X^{01^\prime}=w=x-iy,~X^{10^\prime}={\overline w} = x+iy$ with $\{x,y,z,t\}$ being the 
Cartesian coordinates and the time respectively. Condition (\ref{pfaff}) in 
components reads
\begin{equation}\label{compon}
\frac{\partial S}{~~\partial X^{AA^\prime}}\equiv \partial_{AA^\prime} S=\varphi_A
\psi_{A^\prime}
\end{equation}
and is equivalent to CEE (\ref{eik}) since we have
\begin{equation}\label{det}
\det\Vert \partial_{AA^\prime} S \Vert = \partial_u S\partial_v S - \partial_w S
\partial_{{\overline w}} S = 0,
\end{equation}
because of the factor structure of the r.h.s. of Eq.(\ref{compon}). Conversely, 
the complex gradient 4-vector $\partial_{AA^\prime} S$ should be null in view 
of CEE and, therefore, can be always represented in the form ~(\ref{compon}) with some 
spinors $\varphi,\psi$. The latters are projective in nature being defined up 
to a complex scalar multiplier $\lambda(X)$, 
in view of the following symmetry of Eq.(\ref{pfaff}): 
\beg{symm}
\varphi\mapsto \lambda\varphi,~~\psi\mapsto\lambda^{-1}\psi,~~S\mapsto S.
\eqnd

We proceed now with equivalent transformation of Eq.(\ref{pfaff}) of the form
\begin{multline}\label{change}
dS=\varphi d(X\psi)-(\varphi X)d\psi \equiv \varphi d\beta -\gamma d\psi =\\ 
= \varphi_A d\beta^A - \gamma^{A^\p} d\psi_{A^\p}.
\end{multline}
Here the pair of 2-spinors ${\bf W} = \{\psi,\beta\}\equiv\{\psi, X\psi\}$ related 
to the space-time points (precisely, to the {\it lightcone} of the point $X$) by 
the {\it incidence condition}~\footnote{We have no need to use here the accepted 
multiplication by $i=\sqrt{-1}$ in the r.h.s. of Eqs.(\ref{inc1}). This is allowed under the proper 
redefinition of the standard twistor scalar product, see below}
\begin{equation}\label{inc1}
\beta = X\psi ~~~~(\beta^A = X^{AA^\prime}\psi_{A^\prime})
\end{equation}
forms the {\it projective null twistor} ${\bf W}$. Analogously, the pair of 
2-spinors ${\bf {\overline W}}=\{\varphi,\gamma\}\equiv \{\varphi, \varphi X\}$ forms 
the {\it dual twistor} for which the spinors are related by
\begin{equation}\label{inc2}
\gamma=\varphi X ~~~~(\gamma^{A^\prime} = \varphi_A X^{AA^\prime}).
\end{equation} 
Moreover, twistors ${\bf W,{\overline W}}$ are {\it orthogonal} to each other 
because of the identity
\begin{multline}\label{orthog}
\varphi\beta\equiv\varphi(X\psi)=(\varphi X)\psi\equiv\gamma\psi~~\Rightarrow\\
\Rightarrow ~~<{\bf {\overline W}, W}> = \varphi\beta-\gamma\psi \equiv 0,
\end{multline}
where by $<,>$ the (redifined, see the footnote above) twistor scalar product 
is denoted.

A pair of mutually orthogonal (projective) twistors is known as an 
{\it ambitwistor} ~\cite{lebrun,penrose}. Ambitwistor structure is closely 
related to the flow 
of null geodesics on a generic manifold and, by this, to its Riemannian 
metric structure ~\cite{lebrun}. 
In invariant twistor form CEE (\ref{eik}) can be written as
\begin{equation}\label{compact}
dS=~ <{\bf {\overline W}},d{\bf W}>~ \equiv -<d {\bf {\overline W}},{\bf W}>.  
\end{equation}

Thus, we see that every solution $S(X)$ to CEE may be considered to depend on 
the space-time coordinates $X$ only "implicitly", i.e. via the components of 
projective twistor $W(X)=\{\psi(X),\beta(X)$\} defined by the structure of 
null 4-gradient (\ref{compon}), i.e. $S=S(\psi(X),\beta(X))$. Below, 
it's natural to distinguish two different classes of solutions for which
three components of projective twistor ${\bf W}=\{\psi,\beta\}$ are 
functionally dependent and not.

\section{First class of solutions to CEE}\label{firstclass}

Let us assume that for some solution of CEE three projective components of 
related twistor ${\bf W}\in \mathbb{C}P^3$ are independent with respect to 
space-time coordinates $X$ and in some region of $M$. Making use of the 
symmetry (\ref{symm}), we can consider then that {\it all four twistor 
components} $\{\psi_{0^\prime},\psi_{1^\prime},\beta^0,\beta^1\}$ {\it are 
functionally independent}. Then in view of (\ref{pfaff}) the eikonal 
function $S=S(\psi,\beta)$ will be a function of twistor coordinates while the 
components of the dual twistor $\{\varphi,\gamma\}$ need to have the form
\begin{equation}\label{deriv}
\varphi_A =\frac{\partial S}{\partial \beta^A},~~~ \gamma^{A^\prime}=
-\frac{\partial S}{\partial \psi_{A^\prime}}~.
\end{equation}

Applying now the incidence relation (\ref{inc2}) we get
$$
\frac{\partial S}{\partial \beta^A} X^{AA^\prime} = - \frac{\partial S}
{\partial \psi_{A^\prime}}
$$
and, finally,
\begin{equation}\label{mcond}
P^{A^\prime}(\psi,X)\equiv\frac{dS}{~d\psi_{A^\prime}}\equiv \frac{\partial S}{\partial 
\psi_{A^\prime}} + 
\frac{\partial S}{\partial \beta^A}\frac{\partial \beta^A}{\partial 
\psi_{A^\prime}} = 0.
\end{equation}
 
Two Eqs.(\ref{mcond}) implicitly define two unknown spinor components $\psi_{A^\prime}$ 
at {\it almost every} space-time point $X$ (exept at a subset of zero measure, see 
below). Subsequently resolving algebraic system (\ref{mcond}) for different 
`parameters' $X$ and selecting a continious branch of roots we come to some 
field distribution for the spinor $\psi(X)$ and, accordingly, for $\beta(X)=
X\psi(X)$. Substituting them into $S(\psi,\beta)$ we determine the dependence 
of generating function $S(\psi(X),\beta(X))$ on space-time coordinates.  

To make sure that the function $S(\psi_{A^\prime},X^{AA^\prime} \psi_
{A^\prime})$ which we've obtained really satisfy CEE let differentiate it with 
respect to the coordinates $X^{AA^\prime}$ and get 
\begin{multline}\label{numb}
\partial_{AA^\prime} S =\frac{\partial S}{\partial \psi_{B^\prime}}\partial_
{AA^\prime} \psi_{B^\prime} + \frac{\partial S}{\partial \beta^B}(X^{BB^\prime}
\partial_{AA^\prime} \psi_{B^\prime} + \\ + \delta_{AA^\p}^{BB^\p}\psi_{B^\prime}) 
= P^{B^\prime}(\partial_{AA^\prime}\psi_{B^\prime}) + \frac{\partial S}
{\partial \beta^A} \psi_{A^\prime}.
\end{multline}

In the region of regularity of derivatives $\partial_{AA^\prime} \psi_{B^\prime}$ 
where they all are bounded, we have in account of Eqs.(\ref{mcond}):
\beg{check}
\prt_{AA^\p} S = \frac{\prt S}{\prt \beta_A} \psi_{A^\p}
\equiv \varphi_A \psi_{A^\p}~,
\eqnd
so that the eikonal equation $\det\Vert\prt_{AA^\p} S\Vert = 0$ is
identically satisfied.

On the other hand, the locus where the derivatives of $\psi(X)$ are singular 
corresponds 
to the {\it branching points} of $\mathbb{C}$-valued solutions ({\it multiple 
roots}) of algebraic system (\ref{mcond}), i.e. to the {\it caustic} condition
\beg{sing}
D=\det\Vert\frac{d^2 S}{d\psi_{A^\p}d\psi_{B^\p}}\Vert = 0.
\eqnd

Thus, we have proved the following

\bigskip

\noindent
{\bf Theorem 1.}~ {\it Every (analytical) solution $S(X)$ to CEE for which
all three components of related projective twistor ${\bf W}=\{\psi(X),X\psi(X)\}$ 
are functionally independent, can be generated by some arbitrary (holomorphic) 
twistor function $S(\psi,X\psi)$ via its differentiation and subsequent
resolving of algebraic system (\ref{mcond}) with respect to $\psi_{A^\p}$.
Such solutions have branching points determined by condition (\ref{sing}). 
Conversely, any function $S(\psi, X\psi)$ for which conditions
(\ref{mcond}) can be resolved, define a solution of CEE.} \hfill $\blacksquare$

\bigskip

Of course, the procedure can be simplified by means of the choice of a special 
gauge. Making use of partial arbitrariness of the spinor $\psi$ implied
by the symmetry (\ref{symm}) we can reduce to unity one of the components of 
the spinor $\psi$, say $\psi_{0^\p}$. Then twistor components acquire the form
\beg{twg}
\psi_{0^\p}=1,~~~\psi_{1^\p}\equiv G,~~~\beta^0 = wG+u,~~~\beta^1 = vG+{\overline w},
\eqnd
and generating twistor function reduces to
\beg{gengau}
S = S(\psi^{1^\p},\beta^0,\beta^1) \equiv S(G,wG+u,vG+{\overline w});
\eqnd
it should be then differentiated with respect to $G$, and algebraic equation
\beg{condgau}
\Phi(G,X)\equiv \frac{dS}{dG} = 0,
\eqnd
in the region of regularity, results in some solution $G(X)$ and, subsequently,
-- in some solution $S(G(X),\beta(X))$ of CEE. Singularity condition (\ref{sing})
reduces by this to
\beg{singau}
D=\frac{d^2 S}{dG^2} = 0.
\eqnd
In the form presented by (\ref{gengau}),(\ref{condgau}),(\ref{singau}) the
procedure becomes much similar to that used for determination of singularities
of caustics and wavefronts in the theory of differentiable mappings ~\cite
{arnold,arnold2} and to the method discribed in recent works of E.T. Newman 
with his colleagues ~\cite{newman}. However, twistor methods make the procedure 
quite transparent and preserve Lorentz and gauge invariance of CEE. We'll 
further see also that for $S(X)\in\mathbb{C}$ another class of CEE' solutions 
does exist which has no analogues in the real case. 

Even for the class of solutions dealt with above, unlike the real case where 
always the light-like evolution of wavefronts always takes place, a new type of 
{\it static solutions} (for which both $G(X)$ and $S(X)$
don't depend on time at all) can be obtained in the $\mathbb{C}$-case
since the equation
$$
\vert\vec \nabla S\vert^2 \equiv (\frac{\prt S}{\prt x})^2+(\frac{\prt S}{\prt y})^2     
+(\frac{\prt S}{\prt z})^2 = 0
$$
can then possess nontrivial solutions. As an example, in order to obtain a 
static axisymmetrical solution, let us take the generating twistor function 
\beg{stat}
S=\frac{G^2}{G\beta^0-\beta^1+2iaG}=\frac{G^2}
{wG^2+2 z^* G - {\overline w}}
\eqnd
where constant positive $a\in \mathbb{R}$ and $z^* = z +ia$. Applying the 
procedure described above we easily find 
\beg{statG}
G=\frac{{\overline w}}{z^*}~,
\eqnd
and for CEE' solution correspondent to (\ref{stat})
\beg{statS}
S=\frac{{\overline w}}{{\overline w} w +{ z^*}^2}\equiv\frac{x+iy}{x^2+y^2+(z+ia)^2}.
\eqnd
Both functions (\ref{statG}),(\ref{statG}) are single-valued, and the spinor 
field $G$ is regular on the whole 3-space. However, the eikonal function 
$S$ as well as its complex (null) gradient vector $\vec \nabla S$ {\it become  
singular at a ring of radius} `$a$', i.e. at 
\beg{ring}
z=0,~~~x^2+y^2 = a^2 .
\eqnd
The solution (\ref{statS}) has, therefore, much in common with the famous 
Appel-Kerr static (stationary) solution of the equations of shear-free null
congruence (see below, section 3) whose singular locus is of the same 
ring-like structure as in (\ref{ring}). However, in the case considered we 
have {\it a great number} of static (axisymmetrical) solutions to CEE for 
which the singular locus is bounded in space; we expect to study them in detail,
together with nonstatic solutions, in the forthcoming paper.

\bigskip

\section{\bf Second class of CEE'~solutions and related physical fields}

\bigskip

It remains for us to consider the case when three components of projective 
twistor arising in the form (\ref{change}) of CEE are functionally dependent
\footnote{For a generic twistor, {\it two} independent components do 
always exist, see e.g. ~\cite{joz11}}.  
Making again use of the {\it projective structure} of spinors $\psi,\varphi$ 
manifested by the symmetry 
(\ref{symm}) we can choose the spinor $\psi_{A^\p}$ in such a way that {\it any  
three components of the full twistor ${\bf W}=\{\psi,X\psi\}$ would be functionally 
dependent}.
This means that in the case considered, for every solution of CEE, two 
algebraic constrains between twistor components should exist, of the form
\beg{syst}
\Pi^{(C)}({\bf W}) = \Pi^{(C)}(\psi_{A^\p},X^{AA^\p}\psi_{A^\p}) = 0 
\eqnd
where $\Pi^{(C)},~C=0,1$ are two arbitrary and independent (holomorphic) function 
of four twistor `coordinates'. Analogously to Eqs.(\ref{mcond}) from the previous 
section, algebraic system (\ref{syst}) in implicit form can define, for a pair 
of $\Pi^{(C)}$ being choosed and in the region of regularity, a solution for
the spinor $\psi(X)$.

Let some solution $\psi_{A^\p}(X)$ of Eqs.(\ref{syst}) and, consequently, 
$\beta^A(X)=X^{AA^\p}\psi_{A^\p}(X)$ is found. Then in view of the form (\ref{change}) 
of CEE every (analytical) twistor function $S({\bf W}(X))=S(\psi(X),\beta(X))$ can 
in principle be a solution to CEE (though really it would depend only on two 
functionally independent twistor arguments).

To make it sure we first differentiate with respect to coordinates $X^{AA^\p}$ 
the generating system (\ref{syst}) and get
\beg{der}
\prt_{AA^\p} \psi_{B^\p} =\left [ -\frac{\prt \Pi^{(C)}}{\prt \beta^A} 
Q^{-1}_{(C)B^\p}\right ] \psi_{A^\p} \equiv \Phi_{B^\p A}\psi_{A^\p},
\eqnd
where by $\Phi_{B^\p A}$ the quantities in square brackets are denoted, and 
$Q^{-1}_{(C)B^\p}$ is the matrix inverse of 
\beg{matr}
Q^{(C)B^\p} \equiv \frac{d\Pi^{(C)}}{d\psi_{B^\p}}.
\eqnd

Note that in view of Eqs.(\ref{der}) solutions $\psi_{A^\p}(X)$ of algebraic system 
(\ref{syst}) are defined exept in the singular locus represented by 
the condition
\beg{sing3}
Q\equiv\det\Vert Q^{(C)B^\p}\Vert = \det\Vert\frac{d\Pi^{(C)}}{d\psi_{B^\p}}
\Vert=0.
\eqnd
We'll see further that at these points which correspond to multiple roots 
of Eqs.({\ref{syst}) {\it the strengths of associated physical fields become 
singular}.

Let us now make use of general expression (\ref{numb}) for derivatives of a 
twistor function $S(\psi(X),\beta(X))$. Substituting into it the current 
expression (\ref{der}) for spinor derivatives we get
\beg{der2}
\prt_{AA^\p} S =\left [ -P^{B^\p}\Phi_{B^\p A}+\prt S / \prt \beta^A \right ]
\psi_{A^\p} \equiv \varphi_A \psi_{A^\p}
\eqnd
where $P^{B^\prime}$ are the same as in Eq.(\ref{mcond}) and 
the spinor $\varphi_A$ is now defined by the expression in square brackets. 
We thus see again that the complex 4-gradient $\prt_{AA^\p} S$ is identically 
null because of its factorized structure manifested by the r.h.s. of Eqs.(\ref{der2}), so 
that CEE is satisfied for every generating twistor function. In this way 
we come to the following

\bigskip

\noindent
{\bf Theorem 2.} ~{\it Every (analytical) solution to CEE for which three components 
of related projective twistor ${\bf W}$ are funcionally dependent, can be 
generated by some two arbitrary and independent twistor functions 
$\Pi^{(C)}(\psi,X\psi)$ via resolving the algebraic system (\ref{syst}) with
respect to $\psi(X)$, and is 
represented by arbitrary twistor functions $S(\psi(X),X\psi(X))$ `built' on 
the solutions $\psi(X)$. Branching points for $S(X)$ and $\psi(X)$ are defined 
by the condition (\ref{sing3}).  Conversely, for any pair of $\Pi^{(C)}$
for which the constraints (\ref{syst}) can be resolved, a class of solution of 
CEE $S(\psi(X),X\psi(X))$ is defined.} \hfill $\blacksquare$

\bigskip
   
As in the previous section the procedure can be essentially simplified by
fixing of gauge. Let one of the components of the main spinor, 
say $\psi_{0^\p}$, be nonzero in some region of space-time. Then it can be 
brought to unity and we come to the gauge (\ref{twg}) used before. Now, however, 
three projective twistor components (\ref{twg}) are considered to be functionally 
dependent; the constrains (\ref{syst}) reduce to one algebraic equation of the 
form
\beg{systGG}
\Pi(\psi_{1^\p},\beta^0,\beta^1)\equiv\Pi(G,wG+u,vG+{\overline w}) = 0
\eqnd
which implicitly define an analytical solution $G(X)$ everywhere exept in 
the branching points given by the reduced form of Eqs.(\ref{sing3}):
\beg{sing4}
Q\equiv\frac{d\Pi}{dG} = 0
\eqnd
and, of course, on the curves where $G(X)$ becomes zero or infinite.
Having available a solution $G(X)$ of Eq.(\ref{systGG}) we may be sure that any 
generating function of three projective twistor components `built' on $G(X)$, 
i.e. 
$$
S=S(G(X),wG(X)+u,vG(X)+{\overline w})
$$
(precisely, of any two functionally independent of them) should satisfy CEE. 
For some solutions to CEE $\psi_{0^\p}$ can be zero; in this case another 
gauge can be used to obtain them in analogous manner.

It's worthy of notice that solutions $G(X)$ of generating algebraic equation
(\ref{systGG})  define geometrically the  {\it shear-free null geodesic 
congruences}~\cite{penrose,kerr} which are related, in particular, to 
Riemannian metrics of Kerr-Schild type. Indeed, Eqs.(\ref{der}), in the gauge 
we have used, reduce to the system
\beg{ququ}
\prt_u G = k,~~\prt_w G =kG,~~\prt_{{\overline w}}G = l,~~\prt_v G =l G,
\eqnd
where $k(X),l(X)$  are two remaining nonzero components of the matrix
$\Phi_{B^\p A}$ defined by Eqs.(\ref{der}). Eliminating them from Eqs.(\ref{ququ})
we come to the following nonlinear system:
\beg{ququ1}
\prt_w G = G\prt_u G,~~~~\prt_v G = G\prt_{{\overline w}} G,
\eqnd 
which is known to define the principal spinor of a shear-free null congruence 
and whose {\it general} solution, in accord with the {\it Kerr theorem}
~\cite{penrose,kerr}, is implicitly given by Eq.(\ref{systGG}).

As to the gauge-free representation, Eqs.(\ref{der}) can be rewritten in the 
following invariant matrix form:
\beg{GSE}
d\psi = \Phi dX \psi
\eqnd
($\Phi=\{\Phi_{B^\p A}(X)\}$) which has been earlier proposed in our works 
~\cite{kass,kass2,kass4,joz11}) in the framework of biquaternionic 
generalization of Cauchy- Riemann equations and which was interpreted as a  
basic dynamical system (nonlinear, over-determined, nonLagrangian) for some 
peculiar algebraic field theory ({\it `algebrodynamics'} ). The latter deals
with nontrivially interacting 2-spinor field $\psi(X)$ and $2\times2$-matrix
(complex 4-vector) field $\Phi_{B^\p A}$ and was studied also in
~\cite{joz,joz12,trish}.

Let us briefly enumerate the most important features of this theory 
(for details see ~\cite{joz11,kass11}). For system (\ref{GSE}) an unusual  
gauge symmetry does exist, the so called `weak one', which corresponds to the 
transformations of the form
\beg{weak}
\psi \mapsto \lambda({\bf W})\psi, ~~~\Phi_{B^\p A} \mapsto \Phi_{B^\p A}+
\prt_{AA^\p}\lambda
\eqnd
with a gauge parameter $\lambda({\bf W}) =\lambda(\psi,\beta)$ which 
{\it is allowed to depend on coordinates $X^{AA^\p}$ only implicitly}, via the 
components of the trasforming spinor $\psi(X)$ and of its twistor counterpart 
$\beta(X)=X\psi(X)$. 

Owing to the gradient-wise type of its transformations, the matrix field 
$\Phi_{B^\p A}(X)$ can be interpreted as {\it 4-potentials} of some 
($\mathbb{C}$-valued) `weak' gauge field. Indeed, the integrability 
conditions of Eqs.(\ref{GSE}) $dd\psi\equiv 0$ result in the {\it self-duality of (matrix) field 
strengths} correspondent to potentials $\Phi(X)$ and, in its turn, -- in  
fulfilment of gauge fields' equations. Precisely, {\it both $\mathbb{C}$-valued 
Maxwell and Yang-Mills equations are satisfied identically} on the solutions 
of Eqs.(\ref{GSE}), for the trace and the trace-free part of the full matrix 
field strength respectively. For the 
spinor of complex electromagnetic field strength $F_{(AB)}(X)$ a beatiful 
twistor-like expression was obtained in ~\cite{joz11}:
\beg{emspin}
F_{(AB)} = \frac{1}{Q}\left [ \Pi_{AB} - \frac{d}{dG}\left (
\frac{\Pi_A\Pi_B}{Q}\right ) \right ],
\eqnd
where $\Pi_A,\Pi_{AB},~A,B=0,1$ denote the (1-st and 2-d order) derivatives of 
some function $\Pi(G,\beta^0,\beta^1)$ with respect to its twistor arguments 
$\beta^A$, and $Q=d\Pi/dG$ is the same as in Eq.(\ref{sing4}).
If for any $\Pi$ (after differentiation) the solution $G(X)$ of generating 
Eq.(\ref{systGG}) is substituted into (\ref{emspin}) {\it the spinor 
$F_{(AB)}$ will identically satisfy vacuum Maxwell equations} (exept in the 
points of singularities given by $Q=0$).

Thus, for any solution $G(X)$ of Eq.(\ref{systGG}) condition (\ref{sing4}) 
defines the shape and the time evolution of singularities of some 
associated electromagnetic field. Such singular  
loci can be zero-, one- or two dimensional. In the case when 
the singularities are bounded in 3-space they define some {\it singular,  
particle-like objects} with nontrivial dynamics simulating interaction of 
particles and even their transmutations. 

Moreover, {\it the value of electric charge associated with bounded 
singularities of electromagnetic field (\ref{emspin}) is necessarily quantized} 
(i.e., equal to the whole multiple of some mimimal charge) ~\cite{kass,kass2,kass11}. 
Such ``elementary'' charge is carried, in particular, by the source of the 
exeptional static, axisymmetrical {\it Appel-Kerr solution}
\beg{appel}
G=\frac{{\overline w}}{z^* \pm r}\equiv \frac{x+iy}{(z+ia)\pm \sqrt
{x^2+y^2+(z+ia)^2}}
\eqnd
generated by the function (\ref{systGG}) of the form
\beg{pipi}
\Pi=G\beta^0-\beta^1 = wG^2 + 2 z^* G - {\overline w} = 0.
\eqnd
Its ring-like singularity (\ref{ring}) (branching curve of the $G(X)$) is the 
sameas for the first-class solution of CEE (\ref{statS}); it might be 
interpreted then as a source of electromagnetic field defined by (the real 
part of) Eqs.(\ref{emspin}).  The electromagnetic field itself is
identical to that of the field of the Kerr-Newman solution in GTR} but can carry only `unit', elementary charge ($q=\pm 1/4$ 
in the accepted system of units). General theorem on the charge 
quantization has been proved in ~\cite{kass11}. This fact, 
incidentally, revives the well-known Carter-Lopez model of electron 
~\cite{carter,lopes,joz}. In the limit $a\to 0$ the field turns into the 
`unit-charged' Coulomb field. 

In conclusion, we remind that Eqs.(\ref{GSE}) which we discussed 
above in the framework of general algebraic field theory  are actually  
equivalent to the equations of shear-free null geodesic congruences 
(\ref{ququ1}) and represent the second class of solutions to CEE we studied 
above.

\section{\bf General solution of complex eikonal equation}

We are going now to bring together the results obtained in theorems 1 and
2 and to describe the general solution to CEE (\ref{eik}). 

\bigskip

\noindent
{\bf Theorem 3.}  Let in some open subspace $O\subset M$ there exists an
analytical function $S(X)\in \mathbb{C}$ which is there a solution to CEE.
Then the compoments of projective spinor $\psi_{A^\p}(X)$ and related
projective null twistor ${\bf W(X)}=\{\psi(X),X\psi(X)\}$ are defined via
the structure of null 4-gradient $\prt_{AA^\p} S$, and {\it for every solution 
$S=S({\bf W}$) is a twistor function}, i.e. depends on coordinates $X$ 
only implicitly, as follows:
\beg{twis}
S(X)=S(\psi(X),X\psi(X)).
\eqnd
In (\ref{twis}) the spinor function $\psi(X)$ satisfy either the system 
(\ref{mcond}) or the system (\ref{syst}).

Conversely, any analytical solution to CEE belongs to one of two different
classes and can be generated by one $S(\bf W)$ or two $\Pi^{(C)}, C=0,1$
arbitrary and independent
twistor functions for which {\it algebraic equations (\ref{mcond}) or
(\ref{syst}) can be (analytically) resolved in some subspace $O\subset M$
with respect to $\psi(X)$.} 

In the first case (when $dS/d\psi^{A^\p}=0$), all (projective) twistor
components are independent, and the
solution of CEE is represented by the function $S({\bf W}(X))$ itself, the
branching points being defined by Eq.(\ref{sing}).   

In the second case (when $\Pi^{(C)}=0$), only two projective twistor components
remain functionally 
independent. Then the solution is represented by {\it arbitrary} (analytical)
function $S({\bf W}(X))$ and implicitly depends on $X$ through two
independent twistor arguments only. Branching points are then defined by
Eq.(\ref{sing3}). \hfill $\blacksquare$

To conclude, we show that the Cauchy problem for CEE with arbitrary 
analytical initial data $S(x,y,z,t_0)=S_C$ being given at the moment $t=t_0$ 
can be resolved in a simple algebraic way using the procedure above-presented. 
Indeed, one immegiately obtains at $t=t_0$ the 4-gradient $\prt_{AA^\p} S_C=
\varphi_A\psi_{A^\p}$ and the three components of one of the projective twistors, 
say ${\bf W}=\{\psi, X\psi\}$ (in the matrix $X$ the time coordinate is also 
settled to $t_0$). Now it's a trivial step to find out if these three components 
are functionally independent or not. In the {\it second} case, we immegiately 
obtain (in the gauge for the spinor used) the constrain condition (\ref{systGG}) 
represented by a twistor function $\Pi({\bf W})$. Similarly, in the {\it first} 
case we algebraically express the spatial coordinates $x,y,z$ 
via three {\it independent} twistor components and after substitution into 
$S_C$ come again to a twistor function (\ref{gengau}) $S({\bf W})$. Then, in 
both cases, 
disposing of generating twistor functions $\Pi$ or $S$ and considering now the 
time coordinate in $X$ to be free we are able to obtain the full time-dependent 
solution to CEE corresponding to the given Cauchy data by the prosedure described 
above (i.e. by solving respectively the 
equations $\Pi=0$ or $dS/dG =0$ with respect to the projective spinor 
component $G(X)$). The examples of this quite transparent procedure will be 
presented elsewhere.

\end{document}